\documentclass[twocolumn,showpacs,fleqn,nobibnotes]{revtex4}

\usepackage{amsmath}
\usepackage{graphicx}
\usepackage{float}
\usepackage{subfigure}

\def\lsim{\raise0.3ex\hbox{$<$\kern-0.75em\raise-1.1ex\hbox{$\sim$}}}
\def\gsim{\raise0.3ex\hbox{$>$\kern-0.75em\raise-1.1ex\hbox{$\sim$}}}

\newcommand{\rr}{\mbox{\boldmath $r$}}

\newcommand{\rb}{\mbox{\boldmath $b$}}
\newcommand{\rd}{\mbox{\boldmath $\Delta$}}

\begin{document}

\title{Quarkonium production in coherent hadron-hadron interactions at the LHC}
\author{V.P. Gon\c{c}alves
$^{a}$
and M.V.T. Machado $^{b}$ }

\affiliation{$^a$ Instituto de F\'{\i}sica e Matem\'atica, Universidade Federal de
Pelotas\\
Caixa Postal 354, CEP 96010-900, Pelotas, RS, Brazil.\\
$^b$ Centro de Ci\^encias Exatas e Tecnol\'ogicas, Universidade Federal do Pampa \\ Campus de Bag\'e, Rua Carlos Barbosa,  CEP 96400-700, Bag\'e, RS,  Brazil.}

\begin{abstract}
The photoproduction of quarkonium in coherent hadron-hadron ($pp/pA/AA$) interactions for  LHC energies is an important tool to investigate the QCD dynamics at high energies. In this paper we estimate the integrated cross section and rapidity distribution for $J/\Psi$ and $\Upsilon$ production  using the Color Glass Condensate (CGC) formalism.  We predict large rates, implying that the experimental identification could be feasible at the LHC.

\end{abstract}

\maketitle

\section{Introduction}
In next future the Large Hadron Collider (LHC) at CERN will
start its experimental physics program.  Currently, there is a great
expectation that LHC shall discover the Higgs boson and whatever new
physics beyond the Standard model that may accompany it, such as
supersymmetry or extra dimensions \cite{lhc}. However, several questions remain
open in the Standard Model, which will be probed in a new
kinematical regime at the LHC and determine the background for new
physics. In particular, the description of the high energy regime of the Quantum Chromodynamics (QCD) still is a subject of intense debate (For recent reviews see e.g. Ref. \cite{hdqcd}). 
Theoretically, at high energies (small Bjorken-$x$)  one
expects the transition of the regime described by the linear
dynamics, where only the parton emissions are considered, to a new
regime where the physical process of recombination of partons becomes 
important in the parton cascade and the evolution is given by a
nonlinear evolution equation.  This regime is characterized by the
limitation on the maximum phase-space parton density that can be
reached in the hadron wavefunction (parton saturation), with the
transition being specified  by a typical scale, which is energy
dependent and is called saturation scale $Q_{\mathrm{sat}}$  \cite{hdqcd}.
Experimentally, possible signals of parton saturation have already
been observed both in  $ep$ deep inelastic scattering at HERA and in deuteron-gold 
collisions at RHIC (See, e.g. Ref. \cite{blaizot,universality}). 
 Although the geometrical scaling and the diffractive events observed at HERA, as well as  the high-$p_T$ suppression observed at RHIC, have a natural interpretation in terms of the saturation physics, none of these phenomena can be taken as a conclusive evidence for a new regime of the QCD dynamics. This is due to the kinematical limitations of the experiments. Consequently,  the observation of this new regime still
needs confirmation and so there has bee an active search for new experimental signatures. 

In the last years we have proposed the analysis of coherent hadron-hadron collisions as an alternative way to study the QCD dynamics at high energies \cite{vicmag_upcs,vicmag_hq,vicmag_mesons_per,vicmag_prd,vicmag_pA,vicmag_difper}. The basic idea in coherent  hadron collisions is that
the total cross section for a given process can be factorized in
terms of the equivalent flux of photons of the hadron projectile and
the photon-photon or photon-target production cross section.
The main advantage of using colliding hadrons and nuclear beams for
studying photon induced interactions is the high equivalent photon
energies and luminosities that can be achieved at existing and
future accelerators (For a review see Ref. \cite{upcs}).
 Consequently, studies of $\gamma p$ interactions
at LHC could provide valuable information on the QCD dynamics at
high energies.
The photon-hadron interactions can be divided into exclusive and inclusive reactions. In the first case, a certain particle is produced while the target remains in the ground state (or is only internally excited). On the other hand, in inclusive interactions the particle produced is accompanied by one or more particles from the breakup of the target. The typical examples of these processes are the exclusive vector meson production, described by the process $\gamma h \rightarrow V h$ ($V = \rho, J/\Psi, \Upsilon$), and the inclusive heavy quark production [$\gamma h \rightarrow X Y$ ($X = c\overline{c}, b\overline{b}$)], respectively. Recently, we have discussed both processes considering $pp$ \cite{vicmag_mesons_per,vicmag_prd}, $pA$ \cite{vicmag_pA} and $AA$ \cite{vicmag_hq,vicmag_mesons_per} collisions as an alternative to constrain the QCD dynamics at high energies (For recent reviews see Refs. \cite{vicmag_mpla,vicmag_jpg}). Our results demonstrate that their detection is feasible at the LHC. However, some of these results were obtained considering the  saturation model proposed by Golec-Biernat and W\"{u}sthoff (GBW) several years ago \cite{GBW}. That model has been improved recently considering the current state-of-the-art of  saturation physics: the Color Glass Condensate (CGC) formalism \cite{CGC,BAL,WEIGERT}. This fact motivates a revision  of some our previous estimates. In particular, in this paper  we revise  our predictions for $J/\Psi$ production in coherent $pp/pA/AA$ collisions. Another goal  is to present, for the first time, predictions for the $\Upsilon$ production in coherent collisions considering the CGC formalism.

This paper is organized as follows. In next section (Section \ref{coerente}) we present a brief review of coherent hadron-hadron interactions, introducing the main formulae. In Section \ref{dinamica} we discuss the QCD dynamics and the saturation model used in the calculations. In Section \ref{resultados} we  compare our predictions for the $\gamma p$ cross section with the HERA data and present the  predictions for  quarkonium production in  hadron-hadron interactions. Moreover, we compare the current results to related approaches available in the literature. Finally, in Section \ref{conc} we summarize our main results and conclusions.

\section{Coherent hadron-hadron interactions}
\label{coerente}

Lets consider the hadron-hadron interaction at large impact parameter ($b > R_{h_1} + R_{h_2}$) and at ultra relativistic energies. In this regime we expect the electromagnetic interaction to be dominant.
In  heavy ion colliders, the heavy nuclei give rise to strong electromagnetic fields due to the coherent action of all protons in the nucleus, which can interact with each other. In a similar way, it also occurs when considering ultra relativistic  protons in $pp(\bar{p})$ colliders.
The photon stemming from the electromagnetic field
of one of the two colliding hadrons can interact with one photon of
the other hadron (two-photon process) or can interact directly with the other hadron (photon-hadron
process). One has that  the total
cross section for a given process can be factorized in terms of the equivalent flux of photons of the hadron projectile and  the photon-photon or photon-target production cross section \cite{upcs}. In general, the cross sections for $\gamma h$ interactions are two (or three!) order of magnitude larger than for $\gamma \gamma$ interactions (See e.g. Ref. \cite{vicmag_upcs}). In what follows our main focus shall be in photon - hadron processes.
Considering the requirement that  photoproduction
is not accompanied by hadronic interaction (ultra-peripheral
collision) an analytic approximation for the equivalent photon flux of a nuclei can be calculated, which is given by \cite{upcs}
\begin{eqnarray}
\frac{dN_{\gamma}\,(\omega)}{d\omega}= \frac{2\,Z^2\alpha_{em}}{\pi\,\omega}\, \left[\bar{\eta}\,K_0\,(\bar{\eta})\, K_1\,(\bar{\eta})+ \frac{\bar{\eta}^2}{2}\,{\cal{U}}(\bar{\eta}) \right]\,
\label{fluxint}
\end{eqnarray}
where
 $\omega$ is the photon energy,  $\gamma_L$ is the Lorentz boost  of a single beam and $\eta
= \omega b/\gamma_L$; $K_0(\eta)$ and  $K_1(\eta)$ are the
modified Bessel functions.
Moreover, $\bar{\eta}=\omega\,(R_{h_1} + R_{h_2})/\gamma_L$ and  ${\cal{U}}(\bar{\eta}) = K_1^2\,(\bar{\eta})-  K_0^2\,(\bar{\eta})$.
The Eq. (\ref{fluxint}) will be used in our calculations of quarkonium production in $pA$ and $AA$ collisions. On the other hand, for   proton-proton interactions, we assume that the  photon spectrum is given by  \cite{Dress},
\begin{eqnarray}
\frac{dN_{\gamma}(\omega)}{d\omega} =  \frac{\alpha_{\mathrm{em}}}{2 \pi\, \omega} \left[ 1 + \left(1 -
\frac{2\,\omega}{\sqrt{S_{NN}}}\right)^2 \right] \nonumber \\
\left( \ln{\Omega} - \frac{11}{6} + \frac{3}{\Omega}  - \frac{3}{2 \,\Omega^2} + \frac{1}{3 \,\Omega^3} \right) \,,
\label{eq:photon_spectrum}
\end{eqnarray}
with the notation $\Omega = 1 + [\,(0.71 \,\mathrm{GeV}^2)/Q_{\mathrm{min}}^2\,]$ and $Q_{\mathrm{min}}^2= \omega^2/[\,\gamma_L^2 \,(1-2\,\omega /\sqrt{S_{NN}})\,] \approx (\omega/
\gamma_L)^2$.

The cross section for the photoproduction of a final state $X$ in a coherent  hadron-hadron collision is  given by,
\begin{eqnarray}
\sigma (h_1 h_2 \rightarrow X Y) = \int \limits_{\omega_{min}}^{\infty} d\omega \int dt \,\frac{dN_{\gamma}(\omega)}{d\omega}\,\frac{d\sigma}{dt} \left(W_{\gamma p},t\right)\,,
\label{sigAA}
\end{eqnarray}
where $\frac{d\sigma}{dt}$ is the differential cross section for the process    $(\gamma h \rightarrow X h)$, $\omega_{min}=M_{X}^2/4\gamma_L m_p$, $W_{\gamma p}^2=2\,\omega\sqrt{S_{\mathrm{NN}}}$  and
$\sqrt{S_{\mathrm{NN}}}$ is  the c.m.s energy of the
hadron-hadron system.
Some comments are in order here. Firstly,  the coherence condition limits the photon virtuality to very low values, which implies that for most purposes, they can be considered as real. Moreover, if we consider
$pp/pPb/PbPb$ collisions at LHC, the Lorentz factor  is
$\gamma_L = 7455/4690/2930 $, giving the maximum c.m.s. $\gamma N$ energy
$W_{\gamma p} \approx 8390/1500/950$ GeV. Therefore, while studies of photoproduction at HERA are limited to photon-proton center of mass energies of about 200 GeV, photon-hadron interactions at  LHC can reach one order of magnitude higher on energy. Consequently, studies of coherent interactions at the LHC could provide valuable information on the QCD dynamics at high energies.   Secondly, in $pA$ interactions one have that due to the asymmetry in the collision, with the ion being likely the photon emitter, the photon direction is known, which will implicate an asymmetry in the rapidity distribution (see below).
Finally, in this work we consider that the produced state $X$ represents a quarkonium  ($J/\Psi$ or $\Upsilon$). Since photon emission is coherent over the entire nucleus and the photon is colorless we expect that the events to be characterized by  $Y = h_1 \,h_2$ and two   rapidity gaps.

\section{QCD dynamics at high energies}
\label{dinamica}

Let us consider photon-hadron scattering in the dipole frame, in which most of the energy is
carried by the hadron, while the  photon  has
just enough energy to dissociate into a quark-antiquark pair
before the scattering. In this representation the probing
projectile fluctuates into a
quark-antiquark pair (a dipole) with transverse separation
$\rr$ long after the interaction, which then
scatters off the hadron \cite{nik}.
In the dipole picture the   amplitude for vector meson  production reads  as (See e.g. Refs. \cite{nik,vicmag_mesons,KMW})
\begin{eqnarray}
\, {\cal A}\, (x,\Delta)  = \sum_{n, \bar{n}}
\int dz\, d^2\rr \,\Psi^\gamma_{n, \bar{n}}\,{\cal{A}}_{q\bar{q}}(x,\rr,\Delta) \, \Psi^{V*}_{n, \bar{n}} \, ,
\label{sigmatot}
\end{eqnarray}
where $\Psi^{\gamma}_{h, \bar{h}}(z,\,\rr)$ and $\Psi^{V}_{h,
  \bar{h}}(z,\,\rr)$ are the light-cone wavefunction  of the photon  and of the  vector meson, respectively. The
   quark and antiquark helicities are labeled by $n$ and $\bar{n}$, variable $\rr$ defines the relative transverse
separation of the pair (dipole), $z$ $(1-z)$ is the
longitudinal momentum fractions of the quark (antiquark),  $\Delta$ denotes the transverse momentum lost by the outgoing proton ($t = - \Delta^2$) and $x$ is the Bjorken variable. Moreover, ${\cal{A}}_{q\bar{q}}$ is the elementary amplitude for the scattering of a dipole of size $\rr$ on the target. It is directly related to the $S$-matrix element $S (x,r,b)$ and consequently to the QCD dynamics (see below). One has that \cite{KMW}
\begin{eqnarray}
{\cal{A}}_{q\bar{q}} (x,r,\Delta) & = & i \int d^2 \rb \, e^{-i \rb.\rd}\, 2 [1 - S(x,r,b)]  \nonumber \\
& = & i \int d^2 \rb \, e^{-i \rb.\rd} \,\frac{d \sigma_{q\bar{q}}}{d^2\rb}\,\,,
\end{eqnarray}
where one have introduced the differential dipole-target cross section. Consequently, one can express the amplitude for the photoproduction of a vector meson in the final state as follows
\begin{eqnarray}
 {\cal A}(x,\Delta) = i 
\int dz \, d^2\rr \, d^2\rb  e^{-i[\rb-(1-z)\rr].\rd} 
 (\Psi_{V}^* \Psi_{\gamma})_T \frac{d \sigma_{q\bar{q}}}{d^2\rb}
\label{sigmatot2}
\end{eqnarray}
where $T$ denotes the transverse polarization and one takes into account non-forward corrections to the wave functions \cite{non}.
Finally, the differential cross section  for vector meson photoproduction is given by
\begin{eqnarray}
\frac{d\sigma}{dt} (\gamma h \rightarrow Vh) = \frac{1}{16\pi} |{\cal{A}}(x,\Delta)|^2\,(1 + \beta^2)\,,
\label{totalcs}
\end{eqnarray}
where $\beta$ is the ratio of real to imaginary parts of the scattering
amplitude. For the case of heavy mesons, skewness corrections are quite important and they are also taken  into account. (For details, see Refs. \cite{vicmag_mesons,KMW}). 


The photon wavefunctions appearing in Eq. (\ref{sigmatot2}) are well known in literature \cite{KMW}. For the meson wavefunction, we have considered the Gauss-LC  model \cite{KMW} which is a simplification of the DGKP wavefunctions. The motivation for this choice is its simplicity and the fact that the results are not sensitive to a different model. In photoproduction, this leads only to an  uncertainty  of a few percents in overall normalization. We consider the quark masses $m_c = 1.4$ GeV and $m_b=4.2$ GeV. The parameters for the meson wavefunction can be found in Ref. \cite{KMW} for the $J/\Psi$ case. Accordingly, we have computed the parameters for the $\Upsilon$ case.

The $S$-matrix element $S(x,r,b)$   contains all
information about the target and the strong interaction physics.
In the Color Glass Condensate (CGC)  formalism \cite{CGC,BAL,WEIGERT}, it  encodes all the
information about the
non-linear and quantum effects in the hadron wave function.
It can be obtained by solving an appropriate evolution
equation in the rapidity $y\equiv \ln (1/x)$ and its main properties are: (a) for the interaction of a small dipole ($r
\ll 1/Q_{\mathrm{sat}}$), $S(r) \approx 1$, which characterizes that
this system is weakly interacting; (b) for a large dipole
($r \gg 1/Q_{\mathrm{sat}}$), the system is strongly absorbed which
implies $S(r) \ll 1$.  This property is associate to the
large density of saturated gluons in the hadron wave function. In
our analysis we will  consider the   
saturation model proposed in Ref. \cite{KMW} which generalizes the Iancu-Itakura-Munier (IIM) model \cite{IIM}, introducing the impact parameter dependence.  In this model the differential dipole-proton cross section is parameterized by 
$\frac{d \sigma_{q\bar{q}}}{d^2\rb} = 2 N_p(x,r,b)$, where 
\begin{eqnarray}
N_p(x,r,b) =\,\left\{ \begin{array}{ll}
{\mathcal N}_0 \left(\frac{\bar{\tau}^2}{4}\right)^{\gamma_{\mathrm{eff}}\,(x,\,r)}\,, & \mbox{for $\bar{\tau} \le 2$}\,, \nonumber \\
 1 - \exp \left[ -a\,\ln^2\,(b\,\bar{\tau}) \right]\,,  & \mbox{for $\bar{\tau}  > 2$}\,,
\end{array} \right.
\label{CGCfit}
\end{eqnarray}
$\bar{\tau}=r Q_{\mathrm{sat}}(x,b)$ and the saturation scale is  given by $Q_{\mathrm{sat}}(x,b) = (x_0/x)^{\lambda/2}\,[\exp(-\frac{b^2}{2 B_{CGC}})]^{\frac{1}{2 \gamma_s}}$. The expression for $\bar{\tau} > 2$  (saturation region)   has the correct functional
form, as obtained  from the theory of the Color Glass Condensate (CGC) \cite{CGC}. Moreover,  $\gamma_{\mathrm{eff}}\, (x,\,r)= \gamma_{\mathrm{sat}} + \frac{\ln (2/\tilde{\tau})}{\kappa \,\lambda \,y}$ is the   effective anomalous dimension, which determines the behavior of the dipole cross section in the color transparency regime ($\bar{\tau} < 2$) and introduces evolution effects not included in the GBW model. 
The original parameters of the IIM model were reanalyzed in Ref. \cite{KMW} and are now given by:  $\lambda = 0.159$, $x_0 = 5.95 \times 10^{-4}$, $N_0 =  0.417$, $m_{u/d/s} = 0.14$ GeV and $m_c = 1.4$ GeV. Moreover, we assume in our calculations $B_{CGC} = 5.5$ GeV$^{-2}$. Hereafter, we label this model  by b-CGC. The corresponding dipole-proton cross section is given by $\sigma_{dip}(x,r)=2\int d^2b \,N_p(x,r,b)$.

For  photon-nucleus interactions it is possible to  extend the b-CGC model described above using the Glauber-Gribov formalism  \cite{armesto}. In this case, the nuclear scattering amplitude  $N_A(x,r,b)$  will be parameterized as follows
\begin{eqnarray}
N_A (x,r,b)  = 
\left\{\, 1- \exp \left[-\frac{1}{2} A T_A(b)\, \sigma_{dip}(x,r)  \right] \right\}\,, \nonumber
\label{sigmanuc}
\end{eqnarray}
where $T_A(b)$ is the nuclear profile function, which will
be obtained from a 3-parameter Fermi distribution for the nuclear
density.

\begin{figure}[t]
\includegraphics[scale=0.55]{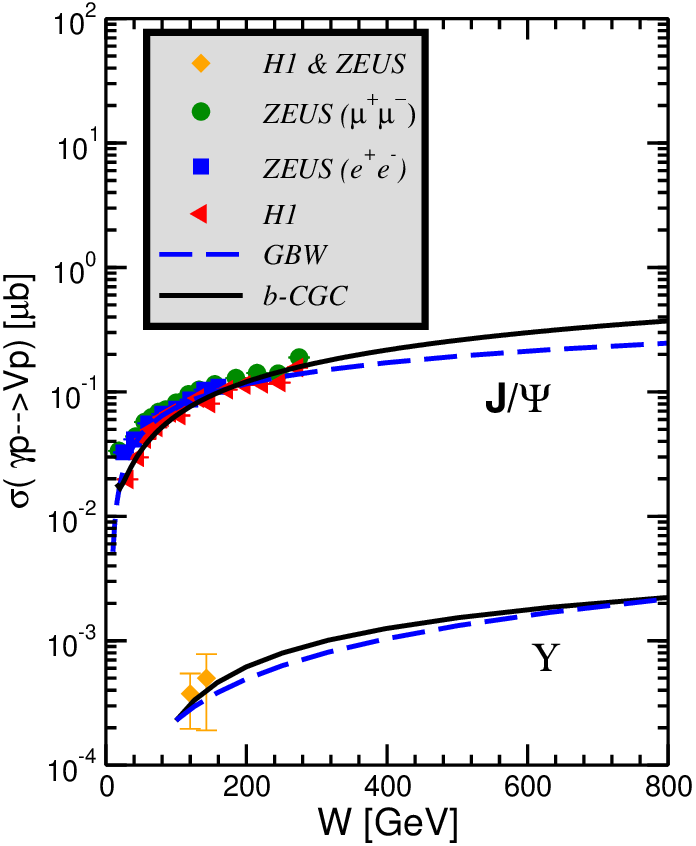}
\caption{(Color online) \it Energy dependence of the $\gamma p$ cross section for quarkonium production. Data from H1 and ZEUS collaborations \cite{H1_jpsi,ZEUS_jpsi}.}
\label{fig:1}
\end{figure}

\section{Results}
\label{resultados}

In Fig. \ref{fig:1} we present the predictions of the GBW and b-CGC models  for the quarkonium photoproduction at HERA and compare it with the current experimental data \cite{H1_jpsi,ZEUS_jpsi}. In particular, it is the first time that the b-CGC model is explicitly compared to the HERA data on vector meson production ($J/\Psi$ and $\Upsilon$). We have that in the HERA kinematical range both predictions are similar and describe the experimental data. However, these are distinct at larger energies, mainly in the $J/\Psi$ case. This result motivates the reanalyzes of our previous predictions for  $J/\Psi$ production in coherent hadron-hadron interactions \cite{vicmag_mesons_per,vicmag_pA}.  Moreover, the description of the $\Upsilon$ production implies that the saturation models are suitable to be used in the calculations of this final state in coherent $pp/pA/AA$ collisions.

\begin{figure}[t]
\includegraphics[scale=0.55]{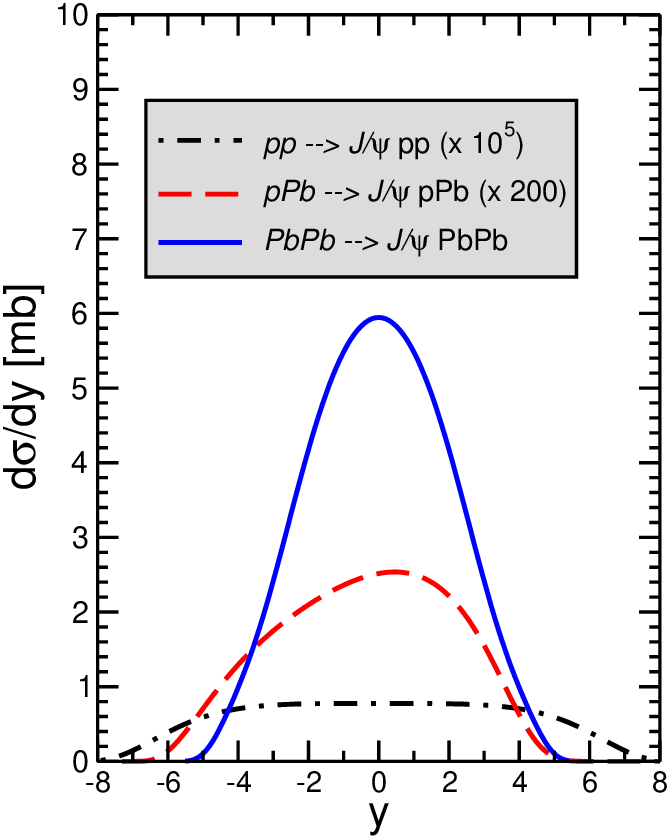}
\caption{(Color online) \it Rapidity distribution for $J/\Psi$ photoproduction on coherent  reactions at the LHC energy (see text).}
\label{fig:2}
\end{figure}

\begin{figure}[t]
\includegraphics[scale=0.55]{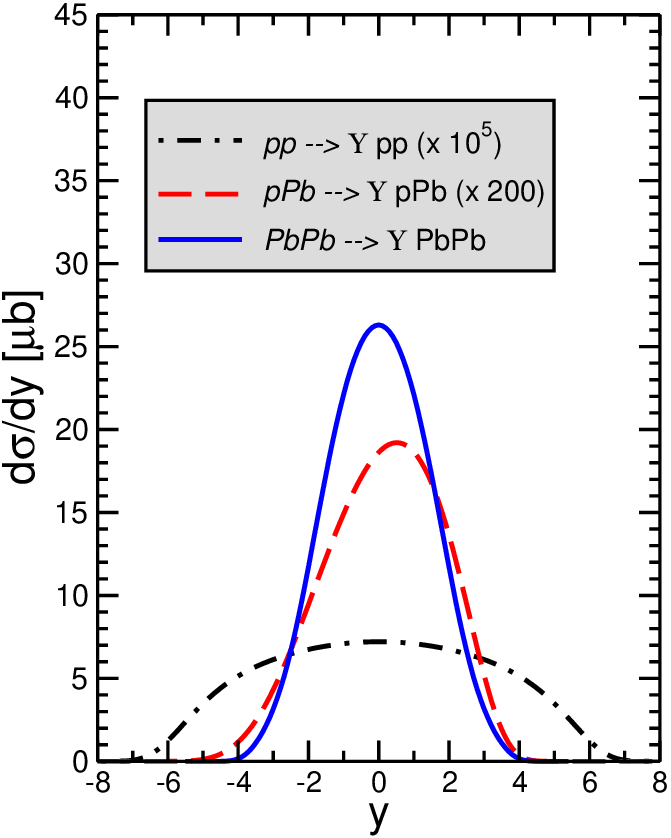}
\caption{(Color online) \it Rapidity distribution for $\Upsilon$ photoproduction on coherent  reactions at the LHC energy (see text).}
\label{fig:3}
\end{figure}

Lets calculate the rapidity distribution and total cross sections for quarkonium production in  coherent hadron-hadron collisions.
The distribution on rapidity $y$ of the produced final state can be directly computed from Eq. (\ref{sigAA}), by using its  relation with the photon energy $\omega$, i.e. $y\propto \ln \, (2 \omega/m_X)$.  Explicitly, the rapidity distribution is written down as,
\begin{eqnarray}
\frac{d\sigma \,\left[h_1 + h_2 \rightarrow   h_1 \otimes X \otimes h_2 \right]}{dy} = \omega \frac{dN_{\gamma} (\omega )}{d\omega }\,\sigma_{\gamma h \rightarrow X h}\left(\omega \right)\,
\label{dsigdy}
\end{eqnarray}
where $\otimes$ represents the presence of a rapidity gap.
Consequently, given the photon flux, the rapidity distribution is thus a direct measure of the photoproduction cross section for a given energy.
In Fig. \ref{fig:2} and \ref{fig:3} we present respectively our predictions for $J/\Psi$ and $\Upsilon$ production in coherent $pp/pA/AA$ collisions, considering $A = Pb$ and $\sqrt{s_{NN}} = 14 / 8.8 /5.5$ TeV, respectively. In the $pp$ case, the production at mid-rapidity at the LHC probes $x$-values of order $(2-7)\times 10^{-4}$. In the $AA$ case, one gets $(6-20)\times 10^{-4}$.

For $pA$ collisions we have an asymmetric rapidity distribution 
since the electromagnetic field surrounding the ion is very larger than the proton one, due to the coherent action of all protons in the nucleus.
As a consequence, the photon direction is known. In contrast, for $pp$ and $AA$ collisions we have symmetric rapidity distributions.  In Table \ref{tabhq} one presents the correspondent integrated cross sections (event rates), using the expected  luminosities. For comparison, we also present our predictions for quarkonium in the ultraperipheral  collision of light nuclei. These results can be contrasted with those obtained for $J/\Psi$ production in Refs. \cite{vicmag_mesons_per,vicmag_pA,klein_prc,klein_nis_prl,strikman_vec}. Firstly, in comparison to our previous estimates for $J/\Psi$ production in coherent $pp$ and $AA$ collisions \cite{vicmag_mesons_per}, we have that the b-CGC predictions are almost identical to those obtained using the GBW model. It is directly associated to the fact that the main contribution for the total cross section comes from low energies, where the predictions of the GBW and b-CGC models for the quarkonium $\gamma p$ production  are very similar.
In comparison to the estimates obtained in Ref. \cite{klein_prc}, our results give higher cross section by a factor of order 11\% for Ca nucleus and are almost similar for Pb nucleus. On the other hand, in comparison with Ref. \cite{strikman_vec} our results are almost 16\% lower for Ca and 41\% lower for Pb. As explained in detail in Ref. \cite{vicmag_mesons_per} the difference between the predictions comes mainly from the distinct QCD approaches used and the different photon flux considered in the calculations. Concerning the $J/\Psi$ production in coherent $pp$ collisions, we have that in comparison to that obtained in Ref. \cite{klein_nis_prl} our prediction is almost 11\% larger.
Finally, for $pA$ collisions, we have that our predictions are almost 12 \% smaller than those obtained in \cite{vicmag_pA}, where we have used the IIM  model \cite{IIM}, which does not consider  the impact parameter dependence of the scattering amplitude.

Lets consider now the $\Upsilon$ production. In this case, our estimate for $pA$ are original in the literature using the CGC approach. In contrast, for $pp$ and $AA$ collision it can be compared with those from Refs. \cite{klein_nis_prl,frank_ups}. In comparison with Ref. \cite{klein_nis_prl} we have that our results for $pp \, (PbPb)$ are approximately a factor 4 (2) smaller. This large difference is directly associated  to the very steep parameterization used in Ref. \cite{klein_nis_prl} for the $\gamma p$ cross section $( \sigma_{\gamma p} \propto W_{\gamma p}^{1.7})$. On the other hand,  our predictions for $\Upsilon$ production in $PbPb$ collisions are approximately 30 \% smaller than those obtained in Ref. \cite{frank_ups} considering the impulse approximation. In comparison with the prediction obtained using the Glauber plus leading twist approximation, our result is approximately 20 \% larger.

\begin{table}[t]
\begin{center}
\begin{tabular} {||c|c|c||}
\hline
\hline
{\bf LHC} & $J/\Psi$ & $\Upsilon$ \\
\hline
\hline
{\bf pp} (${\cal L}=10^{7}$ mb$^{-1}$s$^{-1}$)  & 136 nb (1360.0) & 0.8 nb (8.0) \\
\hline
{\bf pPb} (${\cal L}=420$ mb$^{-1}$s$^{-1}$)  & 85.5 $\mu$b  (36.0) &  0.43 $\mu$b  (0.18) \\
\hline
{\bf CaCa} (${\cal L}=43$ mb$^{-1}$s$^{-1}$)& 440 $\mu$b  (19.0) & 9.7 $\mu$b (0.42) \\
\hline
{\bf PbPb} (${\cal L}=0.42$ mb$^{-1}$s$^{-1}$)& 43.5 mb (18.0) & 0.096 mb (0.04) \\
\hline
\hline
\end{tabular}
\end{center}
\caption{\it The integrated cross section (event rates/second) for the photoproduction of quarkonium in $pp/pA/AA$ collisions at the LHC (see text).}
\label{tabhq}
\end{table}

Finally, two comments concerning the different models used to described the vector meson production and the hadroproduction of vector mesons are in order. Firstly, the vector meson production in general  is described by quite distinct approaches for light and heavy mesons. In the light meson case, the old vector meson dominance (VMD) model is often adopted \cite{klein_prc}, whereas for the heavy mesons a leading logarithmic approximation of the collinear approach is considered \cite{strikman_vec}. In this case, the hard QCD scale is given by the vector meson mass. On the other hand, the dipole picture and saturation physics approach used in this paper is the unique theoretical formalism available that describes simultaneously light and heavy vector meson production \cite{vicmag_mesons}, with the transition between light and heavy mesons being dynamically  introduced by parton saturation effects (via saturation scale) in the  target. Secondly, the vector mesons can also be produced in inclusive and exclusive hadron - hadron interactions. In comparison with inclusive vector meson production, which is characterized by the process $h_1 + h_2 \rightarrow V +X$ (For recent reviews see,
e.g., \cite{hard,lansberg}), one have that the photoproduction cross section is smaller by approximately three order of magnitudes for proton-proton collisions. For nuclear collisions, this factor is smaller due to presence of shadowing corrections. Although the photoproduction cross section would be a small factor of the hadronic cross section, the separation of this channel is feasible if we impose the presence of two rapidity gaps in the final state. This should eliminate almost all of the hadroproduction events while retaining most of the photoproduction interactions. However, two rapidity gaps in the final state can also be generated in exclusive hadron - hadron interactions, characterized by the process $h_1 + h_2 \rightarrow h_1 \otimes V \otimes h_2$. Recently, the exclusive $J/\Psi$ and $\Upsilon$ hadroproduction in $pp/p\bar{p}$ collisions were estimated in Ref. \cite{motyka} considering the pomeron-odderon fusion (See also \cite{schafer}). Although there is a large uncertainty in the predictions for pomeron-odderon fusion, it can be of order of our predictions for the photoproduction of vector mesons. As pointed in Ref. \cite{motyka} the separation of odderon and photon contributions ahould be feasible by the analysis of the outgoing momenta distribution. However, this subject deserves more detailed studies, which we postpone for a future publication.

\section{Summary}
\label{conc}

The QCD dynamics at high energies is of utmost importance for building a realistic description of $pp/pA/AA$ collisions at LHC. In this limit QCD evolution leads to a system with high gluon density. If such system there exists at high energies it can be proven in coherent $pp/pA/AA$ collisions at LHC. In this paper we have analyzed the quarkonium production considering the most recent phenomenological saturation model and demonstrate that it describes the HERA data.  We have investigated the $J/\Psi$ and $\Upsilon$ production in coherent hadron-hadron collisions and showed that the experimental identification could be feasible.

\vspace{1cm}


\begin{acknowledgments}
This work was  partially financed by the Brazilian funding
agencies CNPq and FAPERGS.
\end{acknowledgments}

\end{document}